\documentclass[%
 reprint,
 amsmath,amssymb,
 aps,
]{revtex4-2}

\usepackage{graphicx}
\usepackage{dcolumn}
\usepackage{bm}
\usepackage{amsmath}

\usepackage{float}
\usepackage[nobiblatex]{xurl}
\usepackage{subcaption}
\usepackage{hyperref}

\begin{document}

\preprint{APS/123-QED}

\title{On Punching ``Power":\\Practical Equivalents between improvements\\in Strike Speed and Effective, Striking Mass}

\author{Joseph L. Natale}
\affiliation{
 Fight Science Research Division, Techanique Consulting LLC
}

\date{Sep 12, 2026}

\begin{abstract}
A recent article explores some respective contributions of the initial velocity and effective mass of a striking implement to the ultimately induced, post-collision velocity of a stationary strike target, in a combat sports context. Therein, it was found that multiplicative increases in the effective mass associated with a punch to the human body can have a greater quantitative effect on target velocity than that of its initial, pre-collision speed. Given that fold-increases in effective mass may be limited to low-integer multiples of a modest, literature-supported value that spans martial arts styles -- and the fact that increases in the strike velocity are likely bounded by a factor of two, even for trained fighters -- translating the aforementioned findings into \emph{incremental} changes in either variable may be of more benefit to coaches who intend to apply the theoretical physics of fighting to develop practical training programs. Here, we build on previous work, studying the effects of marginal increases in effective mass on target velocity, and relating  them to
equivalent, single-unit changes in strike speed.
\end{abstract}

\maketitle

\section{\label{intro0}Introduction}

In our recent reconsiderations~\cite{natale_paper0} of the respective roles played by the incoming, effective mass and initial velocity 
in combat-oriented striking, we \emph{i)} derived expressions for the absolute- and fold-changes to the post-collision velocity of a stationary target that result from varying both of these variables, and \emph{ii)} explored limiting cases, within dynamical subspaces that represent realistic human-combat conditions. One key finding from these investigations was that -- despite the ostensible dominance of strike speed $v$ that tends to be espoused by striking coaches, across the martial arts and combat sports milieu -- the effective mass $m$ exerts an effect on the target velocity $V_f$ that is on the same order as
that of $v$, over a considerable range of initial values that  characterize modern sport-fighting.

The above result was discussed generically, in terms multiplicative adjustments to either $v$ or $m$. Here, we fill a theory-to-practice gap by recasting findings from~\cite{natale_paper0} as \emph{marginal value problems} -- wherein we allow the effective mass to vary, then ask \emph{a)} what change in $V_f$ is produced by single-unit (i.e., $1kg$) changes in $m$ from some baseline value, and \emph{b)} what equivalent changes in the strike speed $v$ would be needed to match select, practically-realizable changes in mass, and vice versa, for the previous model of perfectly-elastic strike collisions with a stationary target $M$~\cite{natale_paper0}. This recasting is to facilitate both a more intuitive understanding of the scales over which $m$ and $v$ can be varied, in practice, and the design of specific programs, for training individual athletes whose starting values of these variables, and ultimate needs, themselves vary.

In what follows, we study marginal increases in the value of the effective strike mass from a default, baseline value of $m_b\equiv5kg$, a figure consistent with combat literature~\cite{natale_paper0,lenetsky2015effective}. For this scientific addendum, as before, we begin with analytical results and proceed to a set of illustrative, numerical evaluations. It the author's hope that calculating \emph{incremental} changes and rates-of-return provide further transparency where \emph{multiplicative} increases do not -- and add practical utility by virtue of being more commensurate with progressive-overload training~\cite{tack2013evidence}, for which only small changes can be made consecutively (and wherein the scales of improvements currently thought to be achievable within the combat-sport industry do not far exceed the first few integer multiples of -- or, fold-increase adjustments to -- a $5kg$ baseline in the first place~\cite{natale_paper0}).

\section{Incrementing Mass and Velocity}

We depart from the well-known, ``textbook" result for the post-collision velocity of a stationary target mass $M$ -- struck elastically by a moving mass $m$ -- which we denote by $V_f$. In general, with baseline case $m_b\equiv5kg$ in mind,
\begin{eqnarray}
V_{f}\left(m,M,v\right) ~=~ \left(2\frac{m}{m+M}\right)v
\label{eq:general_Vf}.
\end{eqnarray}

Let $V_{f,b}$ represent the speed picked up by $M$ when $m_b$ is moving with pre-collision velocity of magnitude $v=10\frac{m}{s}$. 
If a single-unit (i.e., $1kg$) increment in effective mass $m$ is applied to the baseline mass, leading to the replacement transformation $m\rightarrow m+1$, simple substitution leads to

\begin{eqnarray}
\Rightarrow  ~V_f\left(m_b+1,M,v\right) ~=~ 2 \cdot\left(\frac{m_b+1}{(m_b+1)+M}\right)v
\label{eq:Vf_plus1kg}.
\end{eqnarray}

More generally, for an arbitrary mass increment $a$, we can take the ratio $q\equiv \frac{V_f\left(m_b+a,M,v\right)}{V_{f,b}}$ in order to write

\begin{eqnarray}
q(m_b,M,a) = \frac{2\frac{(m_b+a)}{(m_b+a)+M}v}{2\frac{m_b}{m_b+M}v} = \frac{\left(m_b+a\right)\left(m_b+M\right)}{m_b\left(m_b+a+M\right)}
\label{eq:q_def},
\end{eqnarray}

by analogy with the factor $k\left(m,M,c\right)\equiv\frac{V_f\left(c\cdot m,M,v\right)}{V_f\left(m,M,v\right)}$ that we introduced in Ref.~\cite{natale_paper0}, for $c\in \lbrace{1,2,3,...\rbrace}$. To gain intuition for the behavior of Eq.~\ref{eq:q_def}, we restore $a=1$, and evaluate $q\left(m_b,M,a\right)$ for the 
baseline value $m_b=5kg$:

\begin{eqnarray}
q(5,M,1) ~=~ \frac{6M+30}{5M+30}
\label{eq:q_m5_a1}.
\end{eqnarray}

The (unitless) expression on the right-hand side of Eq. 4 varies over a characteristically small range, relative to both the  value of the baseline mass in kilograms ($5$) -- or to essentially any $M$ that is likely to be encountered during head-on strikes in human combat; see~\cite{natale_paper0} and Table~\ref{ref_vals_for_combat}). In fact, its limiting value for even supra-physiological values is $\lim_{M\rightarrow\infty} q\left(5,M,1\right)=\frac{6}{5}=1.2$, while its minimum value is $q=1$. In other words, a marginal, $1kg$  increase in the incoming, effective mass results in a  modest, multiplicative change to the post-collision target velocity $V_f$ -- a change which, furthermore, shows diminishing returns as $M$ increases: even for the smaller, $M=m_b=5kg$ target studied in~\cite{natale_paper0}, roughly representative of an opponent's head as the strike target -- $q$ takes on the intermediate value of $\frac{12}{11}\approx1.09$, already $>90\%$ of its maximum.

Even so, these tiny, multiplicative adjustments must, in turn, be multiplied by the strike velocity $v$ in order to treat $q(m,M,a)$ as a proper response function that captures the marginal effect of increasing $m$ from its baseline value. One aspect of Eq.~\ref{eq:general_Vf} that was de-emphasized in~\cite{natale_paper0} (in favor of highlighting the separability of $m$ and $v$), but particularly relevant here, is the inevitability of this ``coupling" between mass- and velocity-increments.

Explicitly, the downstream effect on $V_f$ associated with incrementing $m$, by a fixed amount $a$, becomes stronger not merely when $a$ itself increases -- but also when $v$ is already high. Conversely, the fact that the effect of raising $m$ on $V_f$ weakens as $v$ decreases would seem to imply that incrementing $m$ is less useful than we have established in previous work, for human-scale strike speeds. Still, given that $v$ is bounded by $\sim6\frac{m}{s}$ from below, and practically $v<20\frac{m}{s}$ -- see~\cite{natale_paper0}, and Table~\ref{ref_vals_for_combat} -- the difference between ``low"- and ``high"-velocity contexts for incremental mass augmentations vanishes somewhat, in realistic situations.

Something that remains unaddressed, and yet retains perhaps more practical as well as theoretical value, is the question of how \emph{independent} increases -- in either $m$ or $v$, as the results of specific and distinct training protocols -- can \emph{compensate} for one another. Given a target $M$, what values of $a$ must be realized to produce the same effect on $V_f$ as a marginal increase in the strike velocity? Are these values reasonably within the realm of achievability?

\begin{table}[htbp]
  \centering
  \caption{Reference values for human combat}
  \label{ref_vals_for_combat}
  \begin{tabular}{|l|c|r|} 
    \hline 
    \rule{0pt}{2.5ex}
    Collision Variable & Realistic Range & References \\ [.5ex]
    \hline
     \rule{0pt}{2.5ex}
    Effective Mass $m$ ~& $\sim 5kg$ & \ \cite{neto2007role,walilko2005biomechanics} \\ [.5ex]
    ~Effective Mass $M$ ~ & $5-60kg$ &  \cite{unifiedrules2024} \\ [.5ex]
    ~~(normal leverage)~ &  & \cite{natale_paper0}  \\ [.5ex]
    ~Strike Velocity $v$~ & $6-12\frac{m}{s}$ &  \cite{beranek2023force,corcoran2024impact} \\ [.5ex]
    \hline
  \end{tabular}
\end{table}

\begin{figure}[H]
\includegraphics[width=0.5\textwidth]{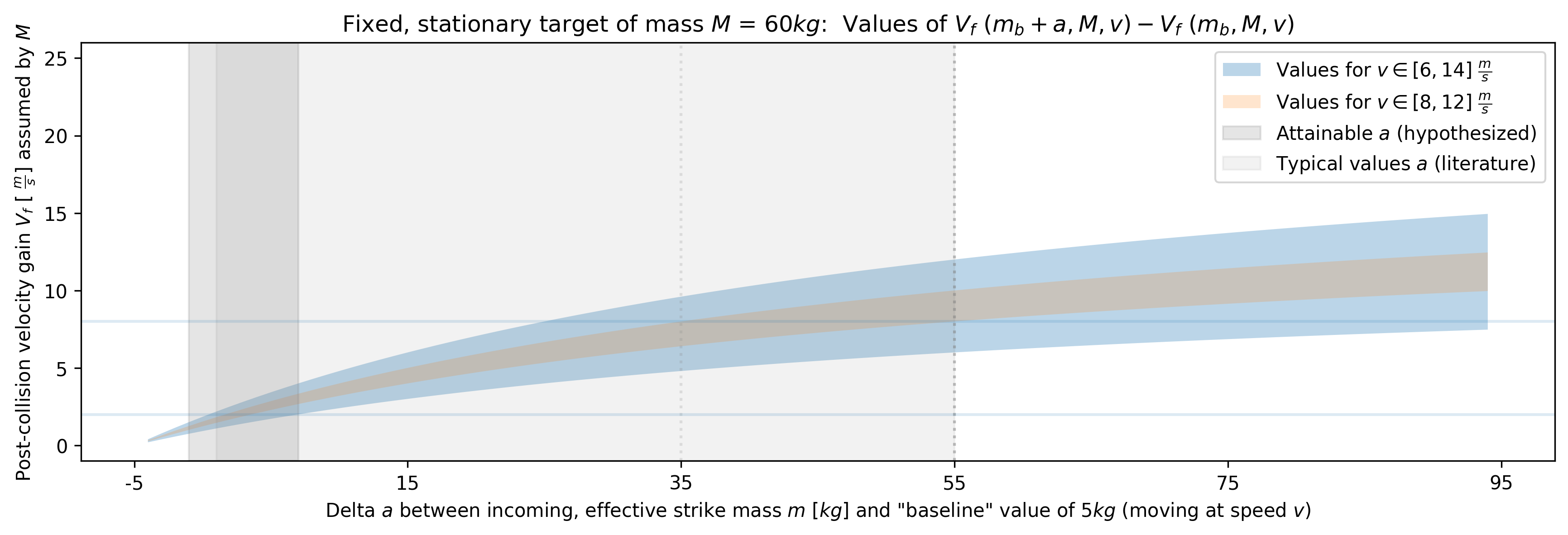}
\caption{\label{fig:vf_shaded_as_prelim_delta} Adapted from Ref.~\cite{natale_paper0}: the post-collision velocity $V_f$ for a stationary target of mass $M=60kg$ grows with increases in both $m$ and $v$. From its value at our decided-upon ``baseline" ($m_b=5kg$) for the incoming, effective mass, the target velocity increases approximately $4.3$-fold once $m$ reaches $30kg$, at any $v\neq 0$; on the other hand, the independent effect of increasing $v$ on $V_f$ is slightly weaker across that $m\in[5,30]$ range; there, the largest difference (occurring at the top of that range, for $m=30kg$), raises $V_f$ twofold, from $4\frac{m}{s}$ (at $v=6\frac{m}{s}$) to $8\frac{m}{s}$ (when $v=12\frac{m}{s}$).}
\end{figure}

Since all ultimate increases in the post-collision, target velocity gains depend linearly on both $q(m,M,a)$ and any multiplicative increases in $v$ -- i.e., $v\rightarrow bv$, for optimistic $1\leq b < 2$ -- we can readily compare the potentially realizable, marginal gains in $V_f$ (above baseline) for achievable values of $q$ with those corresponding to increases in speed $v$, on equal footing. In fact, some order-of-magnitude effects of varying \emph{either} input, on the resulting changes in $V_f$ as output -- at $M=60kg$ -- can be estimated, visually, via Fig.~\ref{fig:vf_shaded_as_prelim_delta}. Next, we isolate the theoretical variables $\Delta m$ and $\Delta v$ to render their effects on $\Delta V_f$ readily calculable.

\section{\label{sec2}Marginal Gain in target velocity}

We have contrasted the face-value independence of $m$ and $v$ (i.e., their separability, in terms of having their own respective, multiplicative influences on $V_f$) with the idea that multiplicative effect of raising either $v$ and $m$ on $V_f$ \emph{intensify at higher starting values of} the other. Based on Fig.~1, we suggest that varying the initial ``strike speed" $v$ (for a fixed value of the effective mass $m$) cannot alter the value of $V_f$ by a significant amount -- on the order of $v$ itself -- until $m$ stands well above $30kg$, or possibly even $60kg$. Beyond this regime, it would be possible to induce fold-changes \emph{and} increases $\sim v$, at least for $M=60kg$.

In particular, once $m\rightarrow M$, $V_f\approx v$, so $V_f$ can \emph{increase} by whatever factor $v$ can improve, within the allowed dynamical ranges associated with human combat~\cite{natale_paper0}. Conversely, since the value of $V_f$ for a baseline mass of $5kg$ is on the order of $1\frac{m}{s}$ -- for example, $V_f(5,60,8)\approx 1.23\frac{m}{s}$, a happy medium between $V_f\approx 0.92\frac{m}{s}$ at the relatively slow $v=6\frac{m}{s}$ and $V_f\approx 1.85\frac{m}{s}$ at the relatively fast $v=12\frac{m}{s}$ -- any opportunities to glean an appreciable increase in $V_f$ \emph{by improving} $v$ \emph{alone} vanish, rapidly, when $m\ll 30kg$.

Converting achievable $V_f$ increases into \emph{marginal gains}, rather than multiplicative, fold-increases or ultimate, raw $V_f$ values, we observe readily that the potential for raising the post-collision velocity of our $M=60kg$ target -- up from the $\sim 1\frac{m}{s}$ order of magnitude corresponding to our baseline incoming, effective mass of $m_b=5kg$ (moving at any speed $v\in[6,12]\frac{m}{s}$) -- is greater for marginal increases in $m$ than it is for $v$, throughout the range $m\in[1,30]kg$.

Furthermore, although marginal contributions of $m$ to increases in $V_f$ -- given by the requisite derivative of Eq.~\ref{eq:general_Vf}, and inferable, via ocular analysis, from Fig. 1 -- vary with the precise value of $m$, this variation is small if $M\gg m$:

\begin{eqnarray}
~\frac{\partial \left(V_{f}\left(m,M,v\right)\right)}{\partial m}  = \frac{2Mv}{\left(m+M\right)^2}
\label{eq:deriv_wrt_m},
\end{eqnarray} 
meaning that there is a factor-of-two decline in the \emph{rate} of accruing marginal gains in $V_f$ for an $M=60kg$ target, even going from a baseline, literature-supported starting value~\cite{lenetsky2015effective} of $m_b=5kg$ to the exaggerated starting value of effective mass $m=30kg$ instead. In general, the highest rate of marginal gains in target velocity $V_f$ is $\approx \frac{2}{M}v$ -- or between $0.2$ - $0.4\frac{m}{s}$ for $M=60kg$, according to the value of $v$, for that first-added kilogram of effective mass; gains slow to $\sim 0\frac{m}{s}$ in additional target speed, per added kilogram of effective mass, when $m\rightarrow\infty$ (i.e., at the point where $V_f\rightarrow 2v)$. Yet, since one recently hypothesized~\cite{natale_paper0}, realizable upper limit for the effective mass of a strike is indeed $\sim 30kg$, or an increase of roughly $25kg$ from the $m=5kg$ baseline, the rate of marginal gains in $V_f$ only drops to a still-appreciable value, about half the magnitude of the gain rate at $m\sim0kg$, even for $m\rightarrow 5m$. This behavior can be compared with the marginal gain rates associated with increasing the strike velocity $v$,
\begin{eqnarray}
\frac{\partial \left(V_f\left(m,M,v\right)\right)}{\partial v}  = \frac{2m}{\left(m+M\right)}
\label{eq:deriv_wrt_v},
\end{eqnarray}

which reach a $V_f$-gain value of $\sim0.15\frac{m}{s}$ per unit (\emph{meter per second}) increase in velocity at the $m_b=5kg$ baseline for a $M=60kg$ target, and do not exceed $\frac{2}{3}\frac{m}{s}$ even when $m=30kg$. Therefore, strategies aimed at increasing $v$ in order to increase $V_f$ require larger starting values of $m$, in order to exert marginal effects on $V_f$ that match what can be achieved with even the first-kilogram addition to the effective mass beyond the baseline, $5kg$ starting point. If it also proves more difficult, in practice, to \emph{continue} to accrue full-unit increments in strike velocity $v$ than it is to add one or \emph{several} kilograms of additionally-contributing body mass to the effective mass $m_b$, training to increase 
speed, exclusively -- at the cost of augmenting $m$ -- would \emph{not} be recommended as a vehicle to effect gains in $V_f$.

All these considerations are subject to change, were our $M=60kg$ target to be swapped for a lighter one, more commensurate with the value of $m$. In the next section, we explore explicit, numerical evaluations for both small- ($m\simeq M$) and large-target ($M=60kg>10m$) scenarios.

\subsection{\label{graphsP1}Cumulative Increases: Speed is inferior to Mass}

In the previous section we first dealt briefly with multiplicative adjustments to the post-collision, target velocity $V_f$; these can be realized by adjusting either the incoming, effective strike mass $m$ or the initial strike velocity $v$. Then, to translate improvement ratios $q(m,M,a)$ into the language of \emph{marginal gains}, or marginal returns on investments in either $m$ or $v$, we used partial derivatives.

Here, we utilize explicit plots to ascertain and visualize how marginal gains in $V_f $ \emph{accumulate} over multiple, unit-increments -- i.e., find the total increase in target velocity, beyond the value $V_{f,b}$ that it would take in the case of a baseline mass $m_b$ moving at speed $10\frac{m}{s}$ -- according to two, related perspectives, when modifying either $m$ or $v$.

First, we trace the family of curves defined by $\Delta V_f  = V_f(m_b+a,M,v) - V_f(m_b,M,v)$, for different values of $v$, as a function of $a$; we highlight the region of values $a=m-m_b$ hypothesized to be broadly accessible, in the context of real combat, according to~\cite{natale_paper0}. This allows us to visualize how potential increases in $V_f$ change if the starting value of mass were on the order of $m_b$, and possibly even smaller, versus an order of magnitude larger. Later, we plot the variation of the derivative $\frac{\partial \left(V_{f}\left(m,M,v\right)\right)}{\partial m}$, as a function of $m$ -- over the corresponding region of interest -- to help determine when increasing the effective mass has the largest, marginal effect on $V_f$.

For the first of the numerical presentations, we produce plots for both a ``large," $M=60kg$ and small, $M=m_b=5kg$, stationary target. Following our examinations of the aforementioned, partial derivative, we ask the question of what the $\Delta V_f$ curves would look like for different starting values of the \emph{incoming}, effective mass (to wit, $m_b>5kg$), choosing an incoming strike speed of $v=10\frac{m}{s}$ 
to answer.

\begin{figure}[H]
\includegraphics[width=0.5\textwidth]{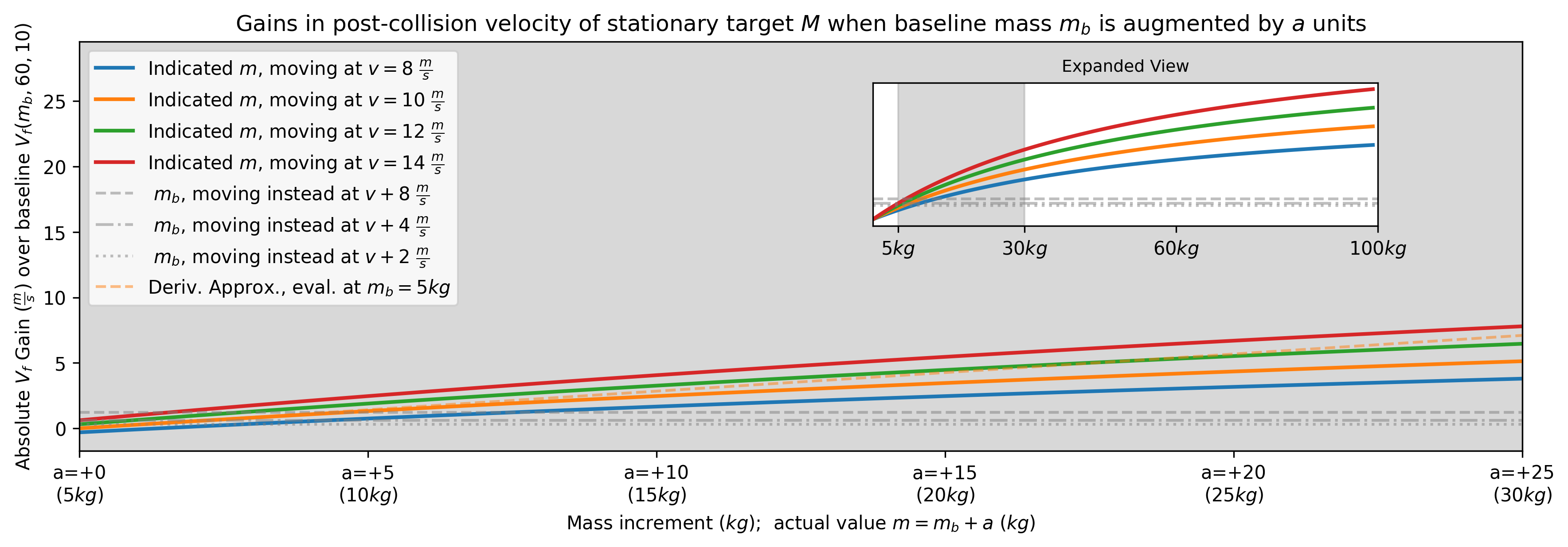}
\caption{\label{fig:vf_gain_with_a__diff_v__60kg} Velocity gain 
(for $M=60kg$) versus the mass-increment $a$. In our region of interest -- pertaining to human combat -- the behavior is similar across different, initial values of the strike velocity, presenting as nearly-linear response curves with slopes that depend only on $v$. Higher $a$ values (see inset) are irrelevant to combat.}
\end{figure}

Turning attention initially to Fig.~\ref{fig:vf_gain_with_a__diff_v__60kg}, 
the velocity gain graph for the $M=60kg$ target, the effect of increasing effective strike mass from its baseline value $m_b$, by a given amount $a$, entails unambiguously positive updates to the corresponding post-collision target velocity, $V_{f,b}\approx.154v$; This effect strengthens when the strike velocity is, itself, increased above its guideline value of $v=10\frac{m}{s}$ (and weakens for slower strike speeds). Indeed, while the nonlinearity in $\Delta V_f$ becomes progressively exaggerated, starting at roughly $m_b+a>4m_b$ -- and coalescing that function into an entirely new, near-constant slope between $m\sim60$ and $m\sim100kg$ -- the highlighted window from $a\in\left[0,25\right]kg$ reveals a nearly linear response to increases $m_b\rightarrow m_b+a$.

This preliminary result -- the apparent linearity of $\Delta V_f$ within our region of interest (values of $m$ that pertain to human combat, on the order of $m_b$, even slightly above) -- should be framed in context of the  behavior of the dotted reference line for the ``mass" derivative (Eq.~\ref{eq:deriv_wrt_m}), evaluated across the horizontal axis: since the definition of \emph{marginal} gain is properly in terms of \emph{single-unit} increments in any given input, multiplying $\frac{\partial V_f}{\partial m}$ 
by the number of unit-mass increments corresponding to any overall increase $a$ serves as an \emph{approximation} of the nominal gain $\Delta V_f = V_f(m_b+a,M,v) - V_{f,b}$. In particular, evaluating the derivative at $m=m_b=5kg$, representing the ``instantaneous" value of the marginal gain in $V_f$ at the baseline mass, precisely, 
shows good agreement with $\Delta V_f$ values from $a=0$ until $a=+5kg$ (i.e., \emph{double} the anticipated effective mass, $m$, for a trained fighter, according to the data in Table~\ref{ref_vals_for_combat}).

Beyond $a=+5kg$ it becomes increasingly difficult, in theory, to increase effective strike mass, via conventional training methods alone; from here until the hypothesized, ``soft" limit around $a=+25kg$~\cite{natale_paper0}, subtle manifestations of nonlinearity --  in this case, a slight ``deceleration," or progressive deficits in the slopes of straight-line tangents, below their initial values, near $a=5kg$ -- cause the actual velocity gain to fall short of those values predicted by the derivative approximation. By $a=+15kg$, na\"ive approximations more closely track the curve for a slightly higher initial strike velocity (e.g., $v=12\frac{m}{s}$ instead of $v=10\frac{m}{s}$).

On the subject of the strike velocity, it is instructive to compare the achievable values of $\Delta V_f$, as defined above, to those which would be attained purely via additive increases $\Delta v$ in speed: $V_f\left(m_b,M,v+\Delta v\right) - V_f\left(m_b,M,v\right)$. To this end, we note that gains in the value of $V_f$, relative to the aforementioned, baseline estimate of $V_{f,b}\approx.154v$, are usurped by effective mass-gain curves, for all velocity increases $\Delta v\in\lbrace 2,4,8 \rbrace \frac{m}{s}$. Such increases in the velocity magnitude are, notably, somewhat nontrivial to achieve in practice -- especially given that the values of $v$ associated with quotidian human combat begin on the order of $\sim10\frac{m}{s}$, per Table~\ref{ref_vals_for_combat} and~\cite{natale_paper0}. Since the marginal gain in $V_f$ with respect to strike velocity remains independent of the velocity \emph{value} itself -- i.e., $\frac{\partial V_f\left(m,M,v\right)}{\partial v}=\frac{\partial V_f}{\partial v}\left(m,M\right)$, as a consequence of Eq.~\ref{eq:deriv_wrt_v} -- the $V_f$-gain curves associated with increases in the striking velocity remain functions \emph{only} of increment $\Delta v$,
and therefore take the forms of straight lines when plotted across the horizontal axis ($a$ values). The intersection of the top, $\Delta v=+8\frac{m}{s}$ line, with mass-increment curve for $v=10\frac{m}{s}$, around $a=+5kg$, implies that even \emph{modest augmentations in the effective mass are equivalent to} \emph{dramatic increases in the strike speed}, when considering realistic values for both $a$ and $m_b$, alike. 

\begin{figure}[H]
\includegraphics[width=0.5\textwidth]{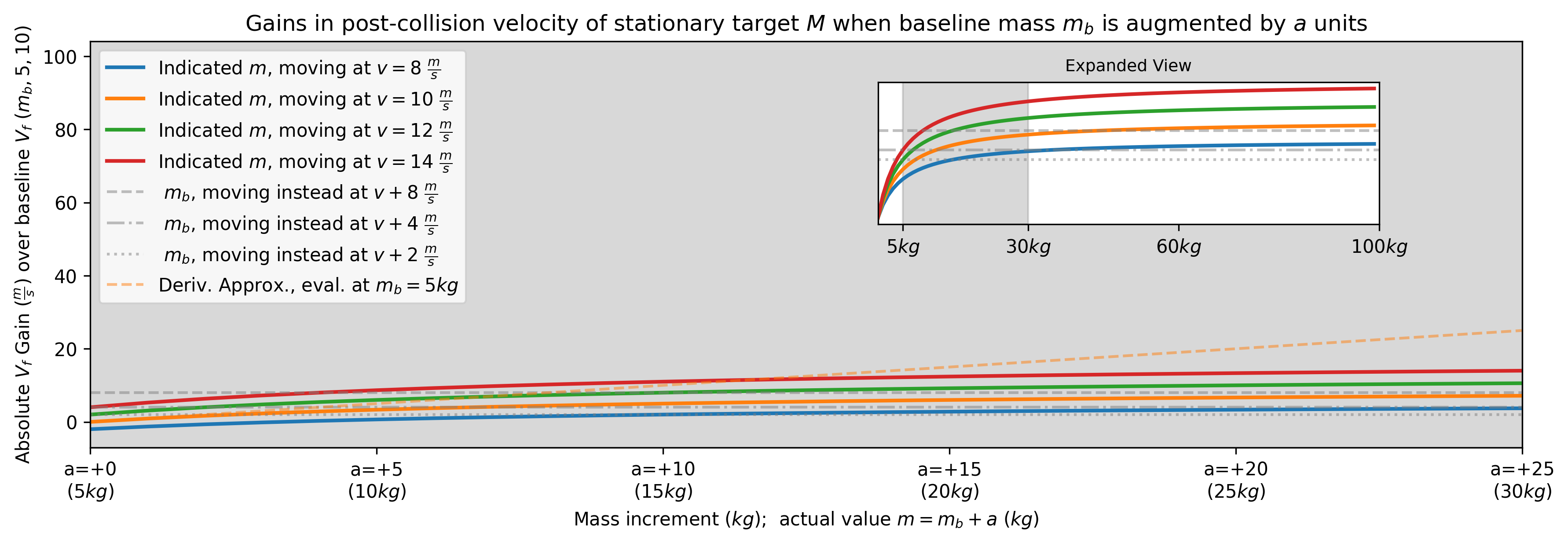}
\caption{\label{fig:vf_gain_with_a__diff_v__5kg} Velocity gain for $M=5kg$, as a function of the mass-increment $a$. Here, the behavior of $\Delta V_f$ is more pronouncedly nonlinear, but separation (absolute difference) of its curves for different initial $v$ is again essentially 
of the same pattern for a given pair of the included strike velocities; 
over half of velocity gains occur for $a\leq 10kg$}.
\end{figure}

In other words, for a heavy target with effective mass $M=60kg$, as long as one considers only effective striking masses on the order of the previously-elucidated~\cite{natale_paper0} $m_b=5kg$ for upper-limb strikes by human combatants, as well as comparably-small increases $a\sim m_b$ (within the immediate and most-likely realm of achievement for most fighters) adding modest mass amounts can yield far more gains in $V_f$ than even an absurd, $80\%$, additive increase in velocity. Although the \emph{marginal} gain at $m_b=5kg$ with respect to mass is narrowly beaten by a full, $\Delta v=+2\frac{m}{s}$ increase in striking velocity (both of these measured relative to $V_{f,b}$; see Section~\ref{sensitivity_to_m} below), this pattern continues to manifest as fundamental even for the smaller-mass ($M=m_b=5kg$) target -- whose corresponding velocity-gain plot can be found in Fig.~\ref{fig:vf_gain_with_a__diff_v__5kg}. Once again, outside the short-lived, initially quasi-linear responses (early slopes, of tangents to the mass-effect curves for low values of $a$), the \emph{velocity}-boosted gain curves intersect the augmented-mass curves at values between $a=+5kg$ and $a=+10kg$.

From this, we infer that -- if they are indeed realizable via technique modifications, without wholly encumbering efforts by any coaches, trainers, or the athlete in question -- additive mass increments, on the order of $a=+5kg$, are superior to training-induced increases in striking velocity, in terms of both higher initial yield (per each anticipated unit of training effort) \emph{and} ultimate growth potential. 

\subsection{\label{graphsP2}Marginal Gains are Highest near Baseline Mass}

So far, we have established that increasing the incoming, effective mass of a modeled strike is (in principle) a more worthwhile avenue than increasing the strike velocity \emph{alone}, 
if the goal is to maximize the gain in the post-collision velocity of a stationary target, \emph{and} an exclusive choice has to be made between these two alternatives -- at least when the starting mass value is in the neighborhood of our decided-upon baseline value, $m_b=5kg$.

It is worth noting that -- although the starting $m$ value could differ from this (arbitrarily-chosen, yet, literature-validated) baseline value, there is a concrete reason why the five-kilogram range is a ``preferred" region for garnering improvements: based on the mathematics of marginal gains, as applied to the physical situation summarized by Eq.~\ref{eq:general_Vf}, the opportunities for improvement in $V_f$ are almost nowhere higher than this modest value when $M\gg m$.

\begin{figure}[htbp]
    \centering
    \begin{subfigure}{\linewidth}
        \centering
        \includegraphics[width=1\textwidth]{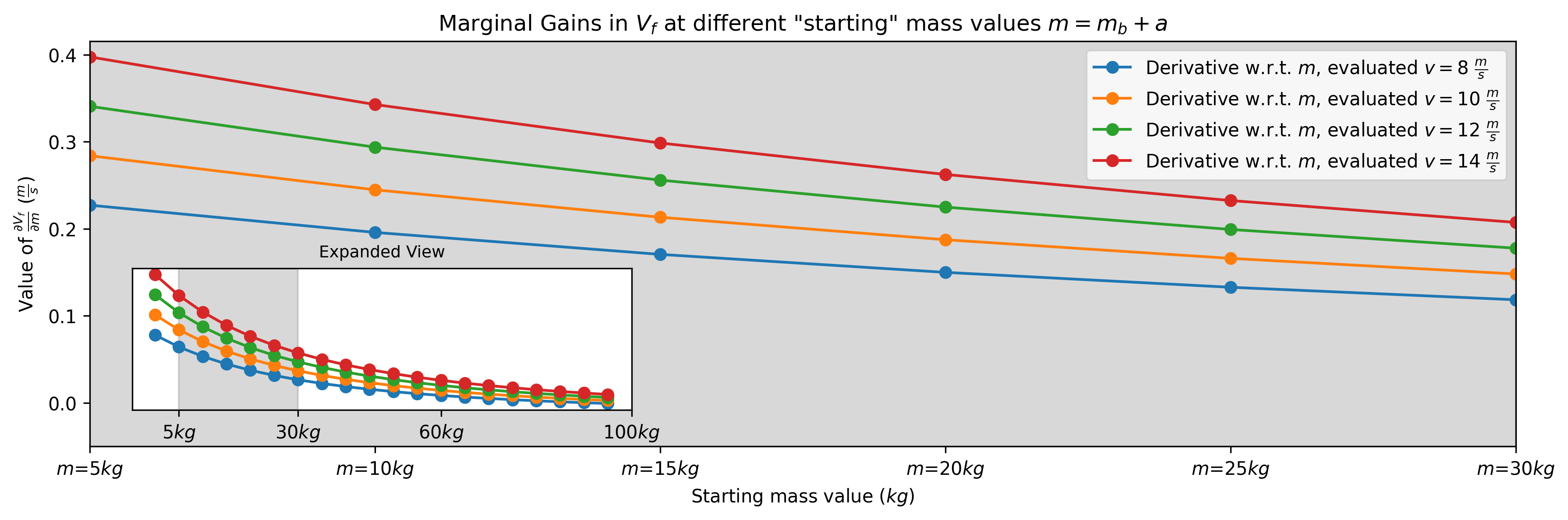}
        \caption{$M=60kg$}
        \label{fig:4a}
    \end{subfigure}
    \vspace{1em}
    \begin{subfigure}{\linewidth}
        \centering
        \includegraphics[width=1\textwidth]{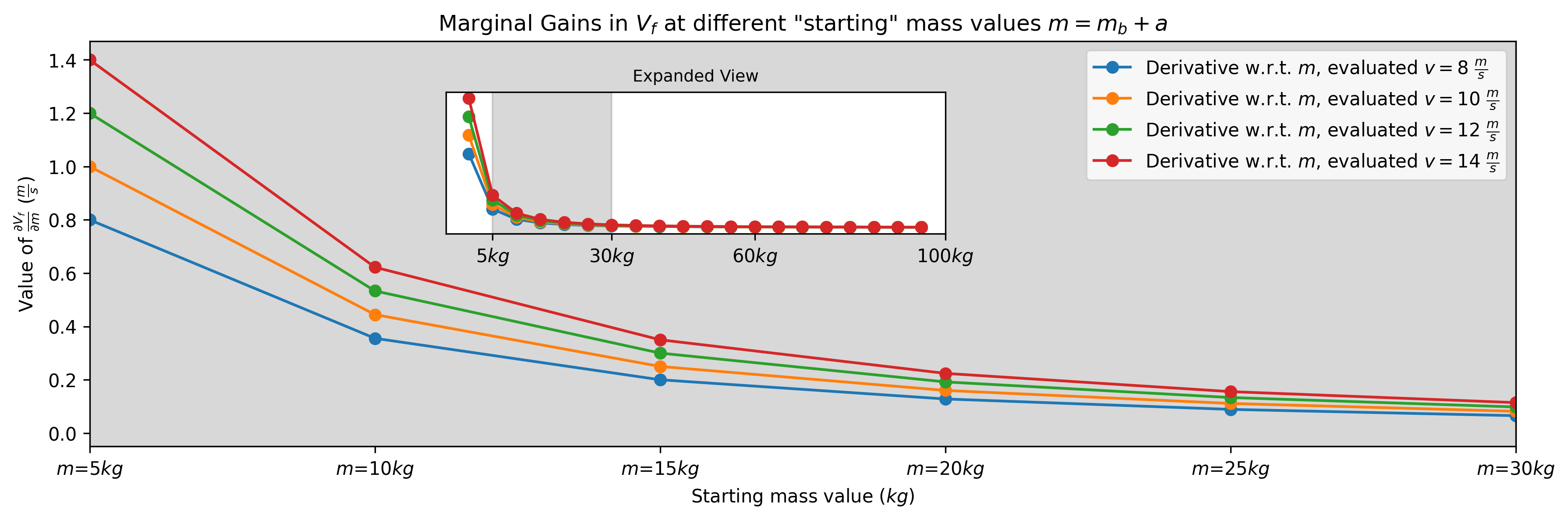}
        \caption{$M=5kg$}
        \label{fig:4b}
    \end{subfigure}
    \caption{Values of the partial derivative $\frac{\partial V_f}{\partial m}$ for different values of $m=m_b+a$, all representing marginal gains in target velocity for $1$-kilogram increases in striking mass.}
    \label{fig:derivatives}
\end{figure}

Below, in Fig.~\ref{fig:derivatives}, we plot values of the derivative $\frac{\partial V_f}{\partial m}$, representing the marginal gain in $V_f$ with respect to unit-increases in the effective mass $m$ (as a function of $m$). It is clear that, irrespective of the velocity value $v$, all marginal gain curves in this graph are monotonically decreasing. This expression of ``diminishing returns" means that the effect of adding a single increment of effective mass ($+1kg$) to a striking implement -- in terms of the corresponding, resulting increase in $V_f$ -- is worth more when the starting value of $m$ is small, and worth less when the $m$ is already very large. This is logically consistent with the constraint that $V_f\rightarrow 2v$, in the limit that $m\rightarrow\infty$; perhaps best understood by describing our elastic collision problem in the ``center-of-mass" frame of reference (not shown here), this limitation  dictates that the $V_f$ \emph{gain} should functionally decrease as $m$ increases.

A second observation is that the ``marginal gain" $\frac{\partial V_f}{\partial m}$, though maximal for minimal $m$, remains nearly as high at $m_b=5kg$ as for just $m=1kg$, when $M=60kg$ (Fig. 4a). While certainly not another near-linear relationship (i.e, considering even the combat-unrealistic upper end), $\frac{\partial V_f}{\partial m}$ decreases steadily (by $\sim50\%$) across the highlighted window. From these facts, it might be inferred that our baseline value of effective mass, $m_b$ -- a value consistent with previous literature aimed at measuring this variable for boxers, mixed martial artists, kung fu practitioners, and the like~\cite{lenetsky2015effective} -- is close to that which affords the \emph{highest possible} $V_f$ gain, given a one-unit increase. Those fighters who ``stand the most to `gain''' are, unsurprisingly, those whose strikes have the smallest, initial $m$ values. 

In other words, the dynamical regime in which even experienced fighters tend to reside, empirically, is precisely the region of \emph{greatest opportunity} to improve the value of $V_f$ via small adjustments ($a$) to any individual's starting value for effective mass $m$ -- at least for large targets $M$.

The diminishing returns associated with the marginal gain in $V_f$ for mass-increases are more pronounced -- that is, the gains decay even more rapidly as a function of the starting mass value than when $M\gg m_b$ -- when $M\sim m_b$. Our evaluations of $\frac{\partial V_f}{\partial m}$ each decrease to around \emph{half} of their values at $m=m_b=5kg$, once the effective striking mass increases to $10kg$, for all included $v$ (Fig. 4b).

\subsection{Sensitivity to starting $m$ depends on target mass\label{sensitivity_to_m}}

In Figure 5, we combine elements of the previous two sections to answer the following question: \emph{what happens to the behavior of the} 
\emph{cumulative gains} in target velocity $V_f$, if the starting value of the effective striking mass $m$ is substituted for something larger than $m_b=5kg$? 
Let us answer, again assuming $v=10\frac{m}{s}$, for a target $M=60kg$.

Beginning with $5$-, $7.5$-, $10$-, and $20kg$ starting masses (i.e., with select, small multiples $c\in\left[1,4\right]$ -- or, increases up to $300\%$ -- of our baseline value $m_b$), we stack cumulative, additive increments, from $a=0$ to $a=+25kg$. Each curve, representing a different starting mass value, therefore begins at the same point, and rises to values on the vertical axis reflecting the cumulative increase in $V_f$ associated with the increments $a$ indicated by the horizontal axis. For the initial increments in effective mass -- until roughly $a=+5kg$ -- none of these curves departs appreciably further than any other. In this regime, thought to be that of a modest and achievable increase in the effective striking mass for many fighters in practice, the post-collision velocity $V_f$ of our stationary, point-mass target $M$ is most sensitive to changes in $m$ (highest marginal gain, given by Eq.~\ref{eq:deriv_wrt_m}; see also Fig.~\ref{fig:derivatives}), so equitable gains $\Delta V_f$ that remain relatively independent of starting mass $m$ dominate the otherwise distinct, cumulative effects. 

\begin{figure}[H]
\includegraphics[width=0.5\textwidth]{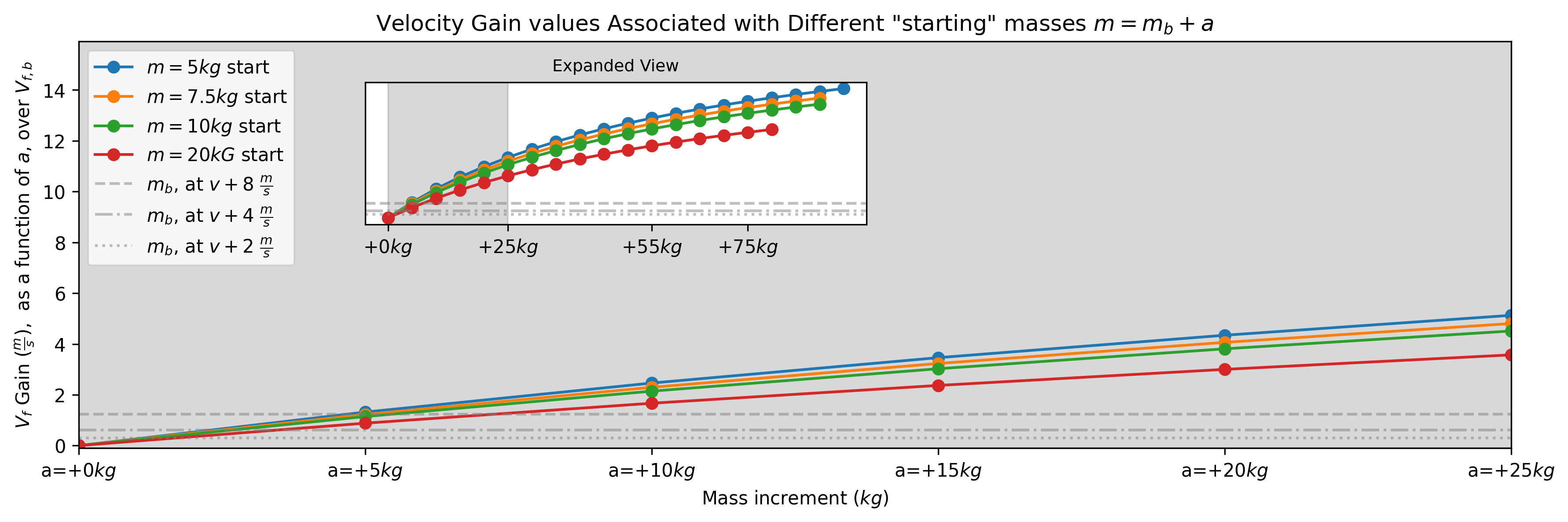}
\caption{\label{fig:delta_Vf_diff_starting_m_60kg} Velocity-gain curves for $M=60kg$ target, with different starting values of the effective striking mass $m$ shown in different colors -- all arranged horizontally, so as to map the corresponding values of $\Delta V_f$ on the vertical axis to different cumulative mass-increments $a$, on equal footing. There is \emph{near}-linearity for a given starting value of $m$, and velocity-gains for $m\in\left[5,10\right]kg$ ``cluster"; only for $m=20kg$ do cumulative increases $a\gtrsim +5kg$ deviate.}
\end{figure}

Following these early, modest increments are those for, roughly, $a\geq 10kg$. In this less-trivially accessible regime -- perhaps most valuable for theoretical purposes, and not representative of practically-achievable increases for most fighters -- it is clear and unsurprising that a starting mass of $m_b=5kg$ is associated with the greatest potential to increase $V_f$: at the risk of a pedantic repetition, we are obliged to acknowledge this consistency with the analyses of the previous section, which explains that fighters with the \emph{smallest} starting $m$ values stand to gain the most, in terms of improving $V_f$ by means of incrementing effective mass alone. Interestingly, though, the \emph{deficit} between the $5kg$ curve and its counterparts -- i.e, any other ``starting" value of the striking mass -- is seen to increase very slowly, on the scale of $\Delta V_f$, for all values of $a$ up to $25kg$. Again, the first few ``extra" mass units are worth approximately the same, when it comes to $\Delta V_f$, regardless of the exact value of the starting mass; even given the special status of $m\lesssim m_b$ (i.e., range theoretically capturing most fighters, especially novices), in terms of ``room for improvement" in $V_f$, essentially anyone can benefit from mass increases, and $a\gtrsim+5kg$ (i.e., $2m_b$) can improve $V_f$ more than $\Delta v$.

To summarize, or reiterate, our last points in the analysis of Fig.~\ref{fig:delta_Vf_diff_starting_m_60kg}, we make two, utilitarian observations. First, based on three supplementary gray lines indicating a hypothetical target's velocity gain due to an increase in \emph{strike speed}, $V_f$ gains that come as a result of incrementing the effective mass continue to dominate after approximately $a=+5kg$ increases, \emph{regardless} of the starting value of mass $m$ -- they are therefore, at least in principle, worth pursuing as a candidate training goal for fighters. Second, even where those plotted mass-increment values $a$, or starting mass values, themselves, lead to $m$ exceeding the expected range proclaimed by Table 1 working with $m=m_b=5kg$ still yields a fair approximation for $\Delta V_f$, provided the effective target mass is large, in comparison with the effective striking mass (i.e., $M\gg m$).

\begin{figure}[H]
\includegraphics[width=0.5\textwidth]{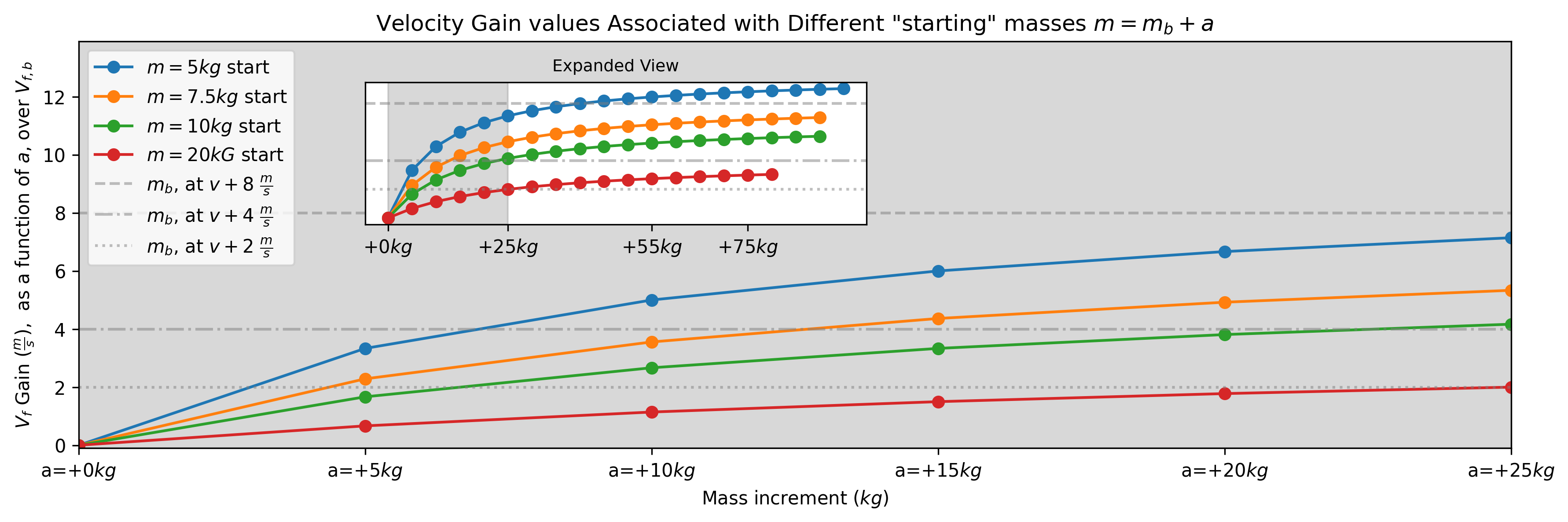}
\caption{\label{fig:vfgain_relative_diffstartingms_5kg} Velocity-gain curves for $M=5kg$ target, with different starting values of the effective striking mass $m$ shown in different colors -- all arranged horizontally, so as to map the corresponding values of $\Delta V_f$ on the vertical axis to different cumulative mass-increments $a$, on equal footing. There are nonlinearities in our ``combat" regime.}
\end{figure}

This relative ``insensitivity" of the cumulative gains in target velocity to the starting $m$ mass \emph{does not} hold for smaller targets that satisfy $M\sim m$. In building Fig.~\ref{fig:vfgain_relative_diffstartingms_5kg} (by substituting $M=m_b=5kg$), we observe that those diminishing returns associated with the potential growth $\Delta V_f$ in target velocity due to increases $a$ in effective strike mass appear in a dramatic way,
when that mass $m$ starts above our chosen baseline value, $m_b=5kg$: if the starting value of $m$ were to reach exactly $20kg$, even the cumulative increases associated with $a=+25kg$ would only generate $\Delta V_f=+2\frac{m}{s}$ -- whereas a gain of this magnitude is surpassed in adding the first $a=+5kg$ of ``extra" mass to the baseline case. Indeed, with the same, exaggerated incrementation in effective striking mass ($a=+25kg$), the $M=5kg$ target would accelerate to $V_f\approx 0.7v$! 

Shrinking returns characterize the individual curves, in  another sense, too -- preserving, in part, the idea we expressed above, that the marginal utility, or incremental output $\Delta V_f$, is greatest for the first, marginal increases in mass, regardless of the precise value $m$ --  starting masses from $m=m_b=5kg$ all the way to $m=20kg$ each realize their maximum contributions to increasing $V_f$ within the early, $a \leq +10kg$ region of the plot for $M=m_b$. Thus, to compare with expectations from the previous sections, it is true \emph{i.)} that the post-collision velocity gains for our stationary target do depend on $M$, in the sense that the smaller-mass target effectively ``spreads" and exaggerates secondary dependencies on the ``starting" value of $m$; \emph{ii.)} that the steepest rise in targets' velocity gains,
relative to baseline $\Delta V_{f,b}$ (nearly half of their maximum, cumulative values for small targets, and perhaps $30\%$ for large targets) are realized within the first $a=+5kg$ of additional mass; \emph{iii.)} that the overall, dynamic range of accessible $\Delta V_f$ values is comparable for $M=5kg$ and $M=60kg$.

\subsection{\label{numbers2}Equivalency in changes to $m$ and $v$} 

Based on the results of the previous sections, it is clear that increasing the initially-moving, 
effective mass associated with a given strike exerts the greatest-magnitude influence on the post-collision velocity of a stationary target $M$ (in terms of facilitating marginal gain in $V_f$) when the value of that mass $m$ lies around the expected~\cite{natale_paper0,lenetsky2015effective} baseline for MMA combatants -- near $m_b=5kg$. In addition, despite the more straightforward, ``zeroth-order" dependence of $\Delta V_f$ on increments to the initial strike velocity (i.e, constant effect, across different starting values of $v$), realistic increments in strike velocity cannot induce $V_f$ gains exceeding those incurred for reasonable, one-to-several-kilogram mass increments $a$ when $m \gtrsim m_b$.

Indeed, it might be informative to ask precisely which increments in mass $a$ are equivalent to specific 
increases $\Delta v$, especially given that strong voices within the combat sports industry have presumed the latter to be the more ``trainable" quality, of these two. In this section we derive expressions to equate these two incrementable variables, in terms of their \emph{marginal} effects on $V_f$ -- as opposed to their \emph{multiplicative} (fold-increase) effects, studied in ~\cite{natale_paper0}.

As an intermediate step toward calculating incremental equivalents, however, we consider the strictly multiplicative relationships elaborated
earlier~\cite{natale_paper0} -- such as the Eq.~\ref{eq:q_def} function $q(m_b,M,a)$ describing the fold-increases between marginally-increased mass configuration $m_b+a$, with output $V_f(m_b+a,M,v)$, and corresponding, starting configuration, with output $V_{f,b}$. Velocity multipliers are more straightforward, taking the form $v\rightarrow bv$, where $b$ is a constant (independent of the striker's effective mass $m$, the effective mass increment $a$, and target mass $M$). 

In general, for any arbitrary effective mass increase of the form above, i.e., $m_b\rightarrow m_b+a$, we know that $$\Delta V_f = V_f(m,M,v) - V_{f,b}(m_b,M,v)$$

or, in terms of the aforementioned, ratio function,
$$\Delta V_f = V_{f,b}(m_b,M,v)\cdot\left(q(m,M,a)-1\right);$$

thus, conversion back to marginal gains ($a=+1kg$) is straightforward. To match the gain accrued via scaling of strike velocity $v$ by its own multiplication factor $b\geq 1$, we must set the product of that constant $b$ with $V_{f,b}$ equal to the product of $q(m,M,a)$ with the same value of $V_{f,b}$. Dropping the baseline target velocity as a common factor, 

\begin{eqnarray}
q_{match} = \frac{\left(a+m_b\right)\left(m_b+M\right)}{m_b\left(a+m_b+M\right)} ~= ~b
\label{eq:79},
\end{eqnarray}

or, substituting in our baseline mass value, $m_b=5kg$,

\begin{eqnarray}
a = \frac{5\left(b-1\right)\left(M+5\right)}{\left(M-5b+5\right)} \
\label{eq:791}.
\end{eqnarray}

Therefore, increasing the initial strike velocity $v$ by an amount $\Delta v = bv-v$, which would, itself, ordinarily lead to $\Delta V_f = V_{f,b} \cdot \left(b-1\right)$, is also, in fact, equivalent to some positive change $a$ in the effective striking mass such that

\vspace{-10pt}
\begin{eqnarray}
\Delta V_f ~= ~  V_{f,b} \cdot \frac{aM}{m_b\left(a+m_b+M\right)}  \
\label{eq:793}.
\end{eqnarray}

As a special case, we might ask for the expected value of $\Delta V_f$ and required, gain-equivalent value of $a$, when the strike velocity is increased by just $+1\frac{m}{s}$ from a reasonable ``starting" value of $v=10\frac{m}{s}$. In this case, $\left(b-1\right)=0.1$, and $\Delta V_f = 0.1V_{f,b}$. For the large, $M=60kg$ target, this modest improvement would be satisfied a mass-increment  (given by the inversion of Eq.~\ref{eq:793}) of $a \approx 0.546kg$: near $m_b$, even a small elevation in mass compensates for a significant change in speed. Even for $M=5kg$, the mass needs only grow by $a=+\frac{10}{9}kg$, or $22\%$ of $m_b$, to compensate.

Below we present a table, for reference, to keep track of several ``equivalent" combinations of $m$, $M$, and $v$. It can be verified that stacking multiple, marginal increments -- for a total, cumulative adjustment $a$ to our baseline mass value $m_b$ -- equates to ``jumping" to the appropriate row (i.e., $m=m_b+a$, with $M$ and $v$ unchanged) in the table.

For large targets, very approximately, each add of $a=+1kg$ to $m_b$ allows a striking implement to travel $\gtrsim 1.5\frac{m}{s}$ slower, while still delivering the baseline impulse $MV_{f,b}$. For both targets, $m_b+1kg$ at $10\frac{m}{s}$ is ``worth" $\Delta v \gtrsim +1\frac{m}{s}$, with the tradeoffs appearing more dramatic for $M=5kg$. Note, in Table~\ref{table2}, that  $\Delta v \sim 3\frac{m}{s}$ can often distinguish a novice from a professional fighter (e.g., within boxing~\cite{kimm2015hand}). This is accomplished just by \emph{doubling} $m_b$ when $M=5kg$.

\begin{table}[htbp]
  \centering
  \caption{Lookup table for ``post-collision" parameters associated with various effective mass (both $m$, striking implement and $M$, target) and combinations with strike velocity (speed) $v$. \textbf{Bold-highlighted} rows indicate the configurations for which post-collision, target speed, $V_f$, is closest (out of all the neighboring rows with the same values of $m$ and $M$) to that of a given $m=6kg$ effective striking mass -- i.e., marginal $a=+1kg$ increase beyond $m_b$ -- moving at $v=10\frac{m}{s}$. This convention applies to an $M=60kg$, and the $M=5kg$ target, respectively, below. 
    \newline \emph{Since $V_f$ is constrained by $\left(0,2v\right]$,
    one needs to consider whether increases $a$, or $\Delta v$, are 
    more readily achievable.}} 
  \label{table2}
  \begin{tabular}{|c|c|c|c|c|}
    \hline
    & & & & \\ [-1em]
    & & & & \\ [-1em]
    Striking Mass & Target & $v$ & $V_f$ & $\Delta V_f-V_{f,b}$\\ [.5ex] \hline

    $m=4kg$ & $M=60kg$ & $8~m/s$ & $1.00~m/s$ & $-0.538~m/s$\\ \hline
    " & " & $10~m/s$ & $1.25~m/s$ &  $+0.288~m/s$\\ \hline
    " & " & $12~m/s$ & $1.50~m/s$ & $+0.038~m/s$\\ \hline
    " & " & $\mathbf{14~m/s}$ & $\mathbf{1.75~m/s}$ & $\mathbf{+0.212~m/s}$\\ \hline

    $m=m_b=5kg$ & " & $8~m/s$ & $1.23~m/s$ & $-0.308~m/s$\\ \hline
    " & " & $10~m/s$ & $1.54~m/s$ & $0~m/s$\\ \hline
    " & " & $\mathbf{12~m/s}$ & $\mathbf{1.85~m/s}$ & $\mathbf{+0.308~m/s}$\\ \hline
    " & " & $14~m/s$ & $2.15~m/s$ & $+0.615~m/s$\\ \hline

    $m=6kg$ & " & $8~m/s$ & $1.45~m/s$ & $-0.084~m/s$\\ \hline
    " & " & $\mathbf{10~m/s}$ & $\mathbf{1.82~m/s}$ & $\mathbf{+0.280~m/s}$\\ \hline
    " & " & $12~m/s$ & $2.18~m/s$ & $+0.643~m/s$\\ \hline
    " & " & $14~m/s$ & $2.55~m/s$ & $+1.01~m/s$\\ \hline

    $m=8kg$ & " & $\mathbf{8~m/s}$ & $\mathbf{1.88~m/s}$ & $\mathbf{+0.344~m/s}$\\ \hline
    " & " & $10~m/s$ & $2.35~m/s$ &  $+0.814~m/s$\\ \hline
    " & " & $12~m/s$ & $2.82~m/s$ &  $+1.29~m/s$\\ \hline
    " & " & $14~m/s$ & $3.29~m/s$ &  $+1.76~m/s$\\ \hline

    $m=2m_b=10kg$ & " & $\mathbf{8~m/s}$ & $\mathbf{2.29~m/s}$ & $\mathbf{+0.747~m/s}$\\ \hline
    " & " & $10~m/s$ & $2.86~m/s$ & $+1.32~m/s$\\ \hline
    " & " & $12~m/s$ & $3.43~m/s$ & $+1.89~m/s$\\ \hline
    " & " & $14~m/s$ & $4.00~m/s$ & $+2.46~m/s$\\ \hline
    
     \hline

    $m=4kg$ & $M=5kg$ & $9~m/s$ & $8.00~m/s$ & $-2.00~m/s$\\ \hline
    " & " & $10~m/s$ & $8.89~m/s$ & $-1.11~m/s$\\ \hline
    " & " & $11~m/s$ & $9.78~m/s$ &  $-0.222~m/s$\\ \hline
    " & " & $\mathbf{12~m/s}$ & $\mathbf{10.7~m/s}$ & $\mathbf{+0.667 m/s}$\\ \hline

    $m=m_b=5kg$ & " & $9~m/s$ & $9.00~m/s$ &  $-1.00~m/s$\\ \hline
    " & " & $10~m/s$ & $10.0~m/s$ & $+0~m/s$\\ \hline
    " & " & $\mathbf{11~m/s}$ & $\mathbf{11.0~m/s}$ & $\mathbf{+1.00~m/s}$\\ \hline
    " & " & $12~m/s$ & $12.0~m/s$ & $+2.00~m/s$\\ \hline

    $m=6kg$ & " & $9~m/s$ & $9.82~m/s$ &  $-0.182~m/s$\\ \hline
    " & " & $\mathbf{10~m/s}$ & $\mathbf{10.9~m/s}$ & $\mathbf{+0.91~m/s}$\\ \hline
    " & " & $11~m/s$ & $12.0~m/s$ &  $+2.00~m/s$\\ \hline
    " & " & $12~m/s$ & $13.1~m/s$ &  $+3.09~m/s$\\ \hline

    $m=8kg$ & " & $\mathbf{9~m/s}$ & $\mathbf{11.1~m/s}$ & $\mathbf{+1.08~m/s}$\\ \hline
    " & " & $10~m/s$ & $12.3~m/s$ & $+2.31~m/s$\\ \hline
    " & " & $11~m/s$ & $13.5~m/s$ & $+3.54~m/s$\\ \hline
    " & " & $12~m/s$ & $14.8~m/s$ & $+4.77~m/s$\\ \hline

    $m=2m_b=10kg$ & " & $\mathbf{9~m/s}$ & $\mathbf{12.0~m/s}$ & $\mathbf{+2.00~m/s}$ \\ \hline
    " & " & $10~m/s$ & $13.3~m/s$ & $+3.33~m/s$\\ \hline
    " & " & $11~m/s$ & $14.7~m/s$ & $+4.67~m/s$\\ \hline
    " & " & $12~m/s$ & $16.0~m/s$ & $+6.00~m/s$\\ \hline
    
  \end{tabular}

\end{table}

\section{Discussion Around Key Results}

We have argued, quantitatively, that increasing the effective striking mass by additive amount $a$ (in kilograms) leads to marginal gains in the post-collision velocity $V_f$ of a stationary target that are on par with, and at times practically exceeding, the corresponding gains in $V_f$ that can be attained through analogous, additive increases -- of realistic magnitude -- in the pre-collision strike velocity.

This is largely \emph{due to the observation} that $v$ only varies by small amounts in practice -- imaginably, $\Delta v \sim 1\frac{m}{s}$ or less, if we stand by the claim that even novice-amateur and professional strikes are only separated by an amount $\Delta v < 4\frac{m}{s}$ -- and, crucially, \emph{predicated on the fact} that the mass-dependent sensitivity of $V_f$ to $v$ only allows major, ``velocity-induced" gains to be achieved when $m$ takes on substantial values, almost an order magnitude above $m_b$. 

We have still not established unambiguously that those ``mass-increment" values $a$ required to rival the $V_f$ gains realizable by modest initial strike velocity $\Delta v$ are, themselves, readily achievable in practice. It was hypothesized previously~\cite{natale_paper0} (based on unpublished, preliminary data) that mass transformations equivalent to some $+300\%$ of $m_b$ are well within reason; it remains to be demonstrated that specific training protocols or technical modifications (i.e., interventions implemented during strike execution) are capable of 
increasing $m$ to such a degree  for already-proficient strikers -- not merely tools for bringing fighters whose effective striking masses start below $m_b$ up to our suggested baseline value~\cite{lenetsky2015effective}. Such strategies are discussed elsewhere~\cite{natale2026_shukjpotentialM}, and do factor into explanations of variation among strikers trained in distinct martial arts styles (as in the ongoing, ``10,000 Fists" study~\cite{natale2026b}), yet remain out of scope for this article. Nevertheless, several comments are in order regarding our supposed values of the effective \emph{target} mass $M$, despite its relative constancy.

Due to the nontriviality of target orientation and shot angle in determining ``real-world" values for the effective mass $M$, both of which are neglected throughout in favor of parsimony, it is likely that earlier estimates~\cite{natale_paper0} of, e.g., $\sim 85\%$ of a full, $70$-kilogram target mass as participating in a given strike collision, \emph{over}-represent the correct value of $M$. It is trivial to calculate that a uniform rod, struck far nearer to its top than its center, might only contribute some $\frac{2}{3}$ of its total mass as a stationary target; only when struck at its \emph{exact} center-of-mass can one even expect $M$ to represent the aforementioned, \emph{total} mass. Accounting for the positional dynamism of any ongoing combat situation, we might  conservatively estimate $M=40kg$, rather than $M=60kg$ as studied throughout, to more faithfully model the ``body shots" first described in~\cite{natale_paper0} -- applied to real, movable, and deformable, material bodies~\cite{albert2017comparison}.

Based on reasoning similar to the argument above for larger body segments, the effective mass of a human head as a target $M$ for intended, ``knockout" strikes might lie closer to $3kg$, rather than being strictly equal to our $m_b \equiv 5kg$. This ``small" target is subject to punishing leverage effects, greatly affecting its effective moment of inertia (e.g., for punches along the chin that result in grand and rapidly-accelerating rotations -- or, even aimed directly, and linearly, at the mid-frontal or frontotemporal region) and therefore any measured values for $M$. In addition, for concussion as one specific, adverse outcome, other physical and physiological nuances complicate injury risk functions in such a way that even bracing the neck muscles (to increase the effective mass $M$ ``seen" by the striking mass $m$) may not 
avoid introducing dangerous $\Delta V_f$, acceleration, and jerk values to brain tissue~\cite{meaney2011biomechanics}.

\bibliography{main}

@PREAMBLE{
 "\providecommand{\noopsort}[1]{}" 
 # "\providecommand{\singleletter}[1]{#1}%" 
}

@unpublished{natale_paper0,
  author       = {Joseph L. Natale},
  title        = {Improving Punching ``Power": Reconsideration of the roles of initial strike velocity and effective mass},
  year         = {2026},
  organization = {Techanique Consulting LLC},
  note         = {Unpublished Manuscript}
}

@article{lenetsky2015effective,
  title={Is effective mass in combat sports punching above its weight?},
  author={Lenetsky, Seth and Nates, Roy J and Brughelli, Matt and Harris, Nigel K},
  journal={Human movement science},
  volume={40},
  pages={89--97},
  year={2015},
  publisher={Elsevier}
}

@article{tack2013evidence,
  title={Evidence-based guidelines for strength and conditioning in mixed martial arts},
  author={Tack, Chris},
  journal={Strength \& Conditioning Journal},
  volume={35},
  number={5},
  pages={79--92},
  year={2013},
  publisher={LWW}
}

@article{neto2007role,
  title={The role of effective mass and hand speed in the performance of kung fu athletes compared with nonpractitioners},
  author={Neto, Osmar Pinto and Magini, Marcio and Saba, Marcelo MF},
  journal={Journal of applied biomechanics},
  volume={23},
  number={2},
  pages={139--148},
  year={2007},
  publisher={Human Kinetics, Inc.}
}

@article{walilko2005biomechanics,
  title={Biomechanics of the head for Olympic boxer punches to the face},
  author={Walilko, Timothy J and Viano, David C and Bir, Cynthia A},
  journal={British journal of sports medicine},
  volume={39},
  number={10},
  pages={710--719},
  year={2005},
  publisher={British Association of Sport and Excercise Medicine}
}

@misc{unifiedrules2024,
  author    = {{Association of Boxing Commissions and Combative Sports}},
  title     = {Unified Rules of Mixed Martial Arts},
  url       = {https://www.abcboxing.com/wp-content/uploads/2024/07/unified-mma-rules-rev-july-2024.pdf},
  year      = {2024},
  note      = {Accessed May 21, 2026}
}

@article{beranek2023force,
  title={Force and velocity of impact during upper limb strikes in combat sports: a systematic review and meta-analysis},
  author={Beranek, Vaclav and Votapek, Petr and Stastny, Petr},
  journal={Sports biomechanics},
  volume={22},
  number={8},
  pages={921--939},
  year={2023},
  publisher={Taylor \& Francis}
}

@article{corcoran2024impact,
  title={Impact force and velocities for kicking strikes in combat sports: A literature review},
  author={Corcoran, Daniel and Climstein, Mike and Whitting, John and Del Vecchio, Luke},
  journal={Sports},
  volume={12},
  number={3},
  pages={74},
  year={2024},
  publisher={MDPI}
}

@article{kimm2015hand,
  title={Hand speed measurements in boxing},
  author={Kimm, Dennis and Thiel, David V},
  journal={Procedia Engineering},
  volume={112},
  pages={502--506},
  year={2015},
  publisher={Elsevier}
}

@unpublished{natale2026_shukjpotentialM,
  author       = {Joseph L. Natale},
  title        = {Investigation of an exception to the ``rules" on the upper limit of effective mass: a proposed mechanism for a new kind of martial arts strike},
  year         = {2026},
  organization = {Techanique Consulting LLC},
  note         = {In preparation.},
}

@unpublished{natale2026b,
  author       = {Joseph L. Natale},
  title        = {The 10,000 Fists Study: Insights from the first 2k Fists},
  year         = {2026},
  note         = {In preparation.}, 
}

@article{meaney2011biomechanics,
  title={Biomechanics of concussion},
  author={Meaney, David F and Smith, Douglas H},
  journal={Clinics in sports medicine},
  volume={30},
  number={1},
  pages={19},
  year={2011}
}

@inproceedings{albert2017comparison,
  title={A comparison of rib structural and material properties from matched whole rib bending and tension coupon tests},
  author={Albert, Devon L and Kang, Yun-Seok and Agnew, Amanda M and Kemper, Andrew R},
  booktitle={IRCOBI Conference Proceedings},
  pages={IRC--17},
  year={2017}
}
\bibliographystyle{unsrtnat}

\end{document}